\documentclass[a4]{elsart5p}

\usepackage{amssymb,amsmath} 
\usepackage{bm,hyphenat,xspace} 
\usepackage{graphicx,epsfig}
\usepackage{tabularx}

\newcommand{\email}[1]{\ead{#1}}

\newcommand {\mcj}{\mathcal{J}}
\newcommand {\mct}{T}

\newcommand{\fm}{\;\mathrm{fm}}

\newcommand{\A}[2]{{}^{#1}\mathrm{#2}}

\newcommand{\Hh}{{}^3\mathrm{H}}

\begin{document}

\begin{frontmatter}

\title {Tetraneutron: Rigorous continuum calculation}

\author{A.~Deltuva}
\email{arnoldas.deltuva@tfai.vu.lt}

\address{
Institute of Theoretical Physics and Astronomy, 
Vilnius University, Saul\.etekio al. 3, LT-10257 Vilnius, Lithuania}



\begin{abstract}
The four-neutron system is studied using exact continuum equations
for transition operators and solving them in the momentum-space
framework. A resonant behavior is found for strongly
enhanced interaction but not a the physical strength, indicating
the absence of an observable tetraneutron resonance,
in contrast to a number of earlier works.
As the transition operators acquire large values at low energies,
it is conjectured that this behavior may explain peaks
in many-body reactions even without a resonance.
\end{abstract}


\end{frontmatter}


\section{Introduction \label{sec:intro}}

The four-neutron ($4n$) system is an exotic few-body system 
challenging experimental techniques as well as theoretical
understanding of the nuclear force and methods for the 
description of the  few-particle continuum.
It has attracted a great interest in the last few years 
\cite{PhysRevLett.116.052501,PhysRevLett.117.182502,PhysRevLett.118.232501,PhysRevLett.119.032501,PhysRevC.93.044004},
but, nevertheless, remains highly controversial. 
An experimental observation of few events in the double 
charge-exchange  reaction $\A{4}{He}(\A{8}{He},\A{8}{Be})$,
that were interpreted as a formation of a tetraneutron resonance
with the energy $E_r = 0.83 \pm 0.65({\rm stat}) \pm 1.25({\rm syst})$ MeV
and  width  $\Gamma \le 2.6$ MeV \cite{PhysRevLett.116.052501},
still awaits a confirmation in the analysis of further experiments.
The theoretical predictions for the tetraneutron are even more contradictory:
They range from a narrow near-threshold resonance
with $E_r \approx 0.8$ MeV and $\Gamma \approx 1.4$ MeV 
\cite{PhysRevLett.117.182502}
or $E_r \approx 2.1$ MeV \cite{PhysRevLett.118.232501}
to a broad resonance with $E_r \approx 7.3$ MeV and $\Gamma \approx 3.7$ MeV 
\cite{PhysRevLett.119.032501} while other authors 
\cite{PhysRevC.93.044004,lazauskas:4n}
predict no observable tetraneutron resonance
at all, i.e.,  negative $E_r$ and very large $\Gamma$.
Despite these differences, all above works 
 concluded that 
tetraneutron properties are insensitive to the details of the 
neutron-neutron ($nn$) and three-neutron ($3n$) interaction models as long as 
they remain realistic. Thus, those very different predictions
cannot be explained by differences in employed potentials but raise
question on the reliability of at least some of the above calculations.
Indeed, the $4n$ system resides in the continuum whose exact 
treatment is much more complicated as compared to bound states.
However, among the above-mentioned works only the solution of the
complex-scaled Faddeev-Yakubovsky (FY) equations
\cite{PhysRevC.93.044004,lazauskas:4n} treats
the continuum rigorously; if no further simplifications
are made unlike in Ref.~ \cite{PhysRevLett.119.032501}
this applies also the no-core Gamow shell model.
In contrast, the harmonic oscillator representation of the continuum 
\cite{PhysRevLett.117.182502} and 
the bound-state quantum Monte Carlo with the extrapolation to the continuum
\cite{PhysRevLett.118.232501}
approaches are not natural methods for
a rigorous description of the four-particle continuum. 
In fact, none of the approaches from
Refs.~\cite{PhysRevLett.117.182502,PhysRevLett.118.232501,PhysRevLett.119.032501}
has been applied successfully to other four-nucleon scattering processes,
in contrast to FY equations \cite{lazauskas:15a}. However,
the only method that so far
provided reliable results for \emph{all} 
four-nucleon reactions above the complete breakup threshold,
i.e., for elastic, charge-exchange, transfer, and breakup processes
in nucleon-trinucleon and deuteron-deuteron collisions,
is the momentum-space transition operator method  
\cite{deltuva:14a,deltuva:15a}.
Furthermore, it provided the most accurate results in the
field of  the universal  four-fermion \cite{deltuva:17d}
and four-boson  \cite{deltuva:12a} physics,
including the properties of resonant (unstable) four-particle states.  
The method is an exact integral version of  FY equations
\cite{yakubovsky:67en}
proposed by Alt, Grassberger, and Sandhas (AGS) \cite{grassberger:67,alt:jinr}.
Its application to the $4n$ problem is highly desirable,
since being a rigorous continuum method it
 should provide  reliable conclusions regarding 
the tetraneutron resonance, much like in the case 
of the $3n$ system  \cite{deltuva:18a},
where it clearly supported the earlier
conclusion \cite{PhysRevC.93.044004,lazauskas:3n}
on the $3n$ resonance unobservability
 in contrast to Ref.~\cite{PhysRevLett.118.232501}.
Another very important advantage of the transition operator
method is its ability to determine not only the resonance position
and width  but also the
nonresonant (background) contribution to scattering
amplitudes, thereby making solid conclusions regarding
the resonance observability in physical processes.

\section{Theory \label{sec:eq}}

AGS equations for four-nucleon transition operators
have been applied to the study of reactions initiated
by all possible two-cluster collisions
 \cite{deltuva:14a,deltuva:15a}. The situation is different
in the $4n$ system that has no bound
subsystems and the only  possible reaction is the
 elastic scattering of four free particles. 
Starting from  Ref.~\cite{fonseca:87},
the operator for this $4\to4$
process can be split into
 two-, three-, and four-particle components, i.e.,
\begin{equation} \label{eq:t44}
T_{4 \to 4} = \sum_j t_j + \sum_{ji\beta} t_j \,  G_0 U_{\beta}^{ji} G_0 \ t_i
+ \sum_{ji\beta\alpha} \mct_{\beta\alpha}^{ji}.
\end{equation}
 Here Latin (sub)superscripts denote
 pairs while Greek subscripts denote two-cluster  partitions
(subsystems) that can be of 3+1 or 2+2 type. Furthermore, 
$G_0$ is the free resolvent at the available energy $E$, 
$t_j = v_j  + v_j G_0 t_j$
are the pair transition operators with pair potentials $v_j$, 
 and  $U_{\beta}^{ji} = G_0^{-1} \bar{\delta}_{ji} +
 \sum_{k} \bar{\delta}_{jk}  t_k G_0  U_\gamma^{ki} $
are the subsystem transition operator where 
$\bar{\delta}_{ji} = 1 - {\delta}_{ji}$. The four-particle transition
operators obey the system of integral equations
\begin{gather}  \label{eq:t4}
\begin{split}
\mct_{\beta\alpha}^{ji} = & {}\sum_k t_j \, G_0 U_{\beta}^{jk} 
\bar{\delta}_{\beta \alpha}  G_0\, t_k\, G_0 U_{\alpha}^{ki} G_0 \, t_i \\ &
+ \sum_{\gamma k} t_j \, G_0 U_{\beta}^{jk} G_0 
\bar{\delta}_{\beta \gamma} \mct_{\gamma \alpha}^{ki}.
\end{split}
\end{gather}
Taking into account identity of neutrons the equations
(\ref{eq:t4}) can be symmetrized, 
reducing the number of $j\beta$ components from 18 to just two,
one being of the 3+1 type and another of the 2+2 type;
in the following they will be abbreviated by subscripts 1 and 2, 
respectively. For example, four-neutron  operators
$\mct_{\beta2}$ are obtained from integral equations
\begin{subequations} \label{eq:t4s}
\begin{align}
 \mct_{12}  = {}&  tG_0 U_1  G_0 t G_0 U_2 G_0 t +
t G_0 { U_1} G_0 ( \mct_{22}- P_{34} \mct_{12}), \\
 \mct_{22}  = {}&  tG_0 U_2 G_0 
 (1  - P_{34})  \mct_{12}, 
\end{align}
\end{subequations}
where $P_{34}$ is the permutation operator of particles 3 and 4,
while $t$ and $U_\beta$ are symmetrized pair and subsystem operators,
respectively \cite{deltuva:07a}.
Kernel of Eqs.~(\ref{eq:t4s}) is built from the same operators
(just in a different order) as in  
Refs.~\cite{deltuva:14a,deltuva:15a,deltuva:07a}
for two-cluster reactions. Thus, the solution technique to a large extent
can be taken over from 
Refs.~\cite{deltuva:14a,deltuva:15a,deltuva:07a}. It is performed in the
momentum-space partial-wave representation
\cite{deltuva:07a}, whereas kernel singularities
arising from $G_0$ are treated by the complex-energy
method with special integration weights \cite{deltuva:12c}.
As the four-cluster matrix elements exhibit stronger
dependence on the imaginary part $\varepsilon$ of the energy,
smaller values $0.1$ {MeV} $  \le \varepsilon \le 1$ MeV
as compared to Refs.~\cite{deltuva:14a,deltuva:15a,deltuva:12c}  
have to be used, which implies larger number (around 80)
of grid points for the discretization of Jacobi momenta
$k_x$, $k_y$, and $k_z$ in the notation of 
Refs.~\cite{deltuva:12a,deltuva:12c}.

A pure $4n$ scattering experiment is  practically impossible,
with presently available experiment techniques
one may only indirectly observe $4n$ as a final subsystem in a 
more complicated reaction such as 
 $\A{4}{He}(\A{8}{He},\A{8}{Be})$. It is complicated many-body
process that cannot be described rigorously by presently
available methods, however,  half-shell 
matrix elements of $\mct_{\beta\alpha}$ that determine the $4n$ 
wave function, together with
some simplified reaction model,  may provide estimation
for the properties of the  final $4n$ subsystem,
e.g., its energy distribution. Therefore it is important
to evaluate also half-shell matrix elements of $\mct_{\beta\alpha}$.

\section{Results \label{sec:res}}

In the following I consider the $4n$ state with total angular
momentum and parity $\mcj^\Pi = 0^+$; namely in this state Refs.
\cite{PhysRevLett.117.182502,PhysRevLett.118.232501,PhysRevLett.119.032501}
predict the $4n$ resonance. In order to make comparison with those
works, I use chiral effective field theory ($\chi$EFT) 
potential at next-to-leading
 order (NLO) ~\cite{PhysRevLett.115.122301},
an improved version of the local NLO potential
used in Ref.~\cite{PhysRevLett.118.232501}, and a low-momentum
potential that should have similar behavior as those
used in Refs.~\cite{PhysRevLett.117.182502,PhysRevLett.119.032501}.
 It is based on
a realistic Argonne V18 potential \cite{wiringa:95a} evolved
using the similarity 
renormalization group (SRG) transformation \cite{bogner:07c} 
with the flow parameter $\lambda = 1.8\, \fm^{-1}$. It is important
that this is one of  few models able to reproduce quite 
well not only the $\Hh$ binding energy but also the cross
section for $n$-$\Hh$ scattering in the energy regime 
with pronounced four-nucleon ($3n$ + proton) resonances
\cite{deltuva:08b}. For this reason it can be considered 
as a well suited model for the  $4n$ resonance study.

In order to follow the evolution of the  $4n$ resonance,
I also perform calculations enhancing the $nn$ potential 
by a factor $f>1$ in $nn$ partial waves with the total angular
momentum $j_x < 3$ except for the  ${}^1S_0$ wave where
the physical potential strength is kept, ensuring that 
there are no bound dineutrons.  
The calculations include $nn$ partial waves with $j_x < 3$
and $3n$ partial waves with the total angular
momentum $J_y < \frac72$, while subsystem
orbital angular momenta are $l_y,l_z < 5$.
With these cutoffs the results appear to be well converged.

Since the $4n$ resonance corresponds to the pole of the
transition operators $\mct_{\beta\alpha}$ in the complex
energy plain at $E_r - i\Gamma/2$, these values are extracted 
from the energy dependence of calculated
$\mct_{\beta\alpha}$ matrix elements   much like 
in Ref.~\cite{deltuva:18a} for the $3n$ resonance.
In general they  are functions of six Jacobi momentum variables,
but since all of them exhibit the same resonant behavior,
they are shown for few initial and final on-shell and off-shell states only,
abbreviated by $|k_a\rangle$ and $|k_a^{\rm off}\rangle$, respectively.
They are chosen with $l_x=l_y=l_z=0$ and two vanishing Jacobi momenta
$k_j =0$ for $j\ne a$, the remaining one being
 $k_a = \sqrt{2\mu_a E}$ and 
$k_a^{\rm off} = \sqrt{2\mu_a (E+\epsilon_{\rm off})}$, respectively.
For example, the state $|k_z\rangle$ in the 2+2 configuration
corresponds to two pairs of neutrons with vanishing relative $nn$ momentum,
that can be interpreted as a two (unbound) dineutron state.
In the above relations $\mu_a$ is the associated reduced mass while 
$\epsilon_{\rm off}$ measures how much off-shell the system is.
A typical value in the shown results is $\epsilon_{\rm off} = 2$ MeV
that roughly  corresponds to $\A{8}{He}$ binding
with respect to the $\A{4}{He}+4n$ threshold.

\begin{figure}[!]
\begin{center}
\includegraphics[scale=0.66]{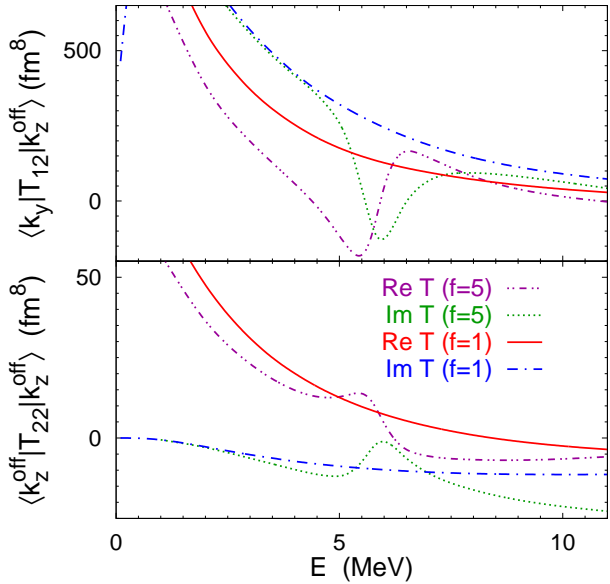}
\end{center}
\caption{\label{fig:treim} (Color online)
Energy dependence of real and imaginary parts
of  selected $\mcj^\Pi = 0^+$
four-neutron transition matrix elements
calculated using the SRG potential with higher wave
 enhancement factors $f=1$ and 5.}
\end{figure}

\begin{figure}[!]
\begin{center}
\includegraphics[scale=0.64]{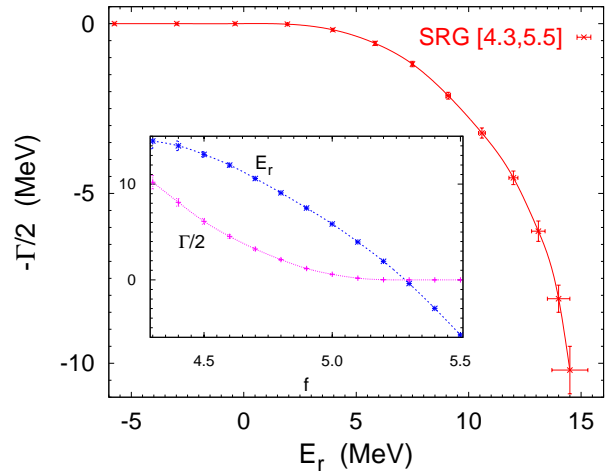}
\end{center}
\caption{\label{fig:eg} (Color online)
Four-neutron  $\mcj^\Pi =0^+$ resonance trajectory  obtained with the
 SRG potential varying 
the higher-wave enhancement factor $f$ from 5.5 to 4.3  with the step of 0.1.
The inset shows the individual dependence of $E_r$ and $\Gamma$ on f.
Lines are for guiding the eye only.}
\end{figure}

 $4n$ transition operators $\mct_{\beta\alpha}$ in the $\mcj^\Pi = 0^+$ wave
calculated using the physical $nn$ potential, i.e., $f=1$, 
show no indications of resonance, but for sufficiently large
$f$ a resonant behavior is clearly seen in all  matrix elements;
two examples for the SRG potential with $f=1$ and  $f=5$ are presented in 
Fig.~\ref{fig:treim}. The results indicate that
nonresonant contributions are very important
 even at  $f=5$ with $E_r - i\Gamma/2 \approx (5.9-0.6i)$ MeV.
The  $\mcj^\Pi = 0^+$ resonance position and width extracted
at different $f$ values are displayed in Fig.~\ref{fig:eg}.
The $4n$ system becomes bound at $f=5.29$. Thus, the tetraneutron
is lower in energy than the trineutron that in the SRG model becomes 
bound only at  $f>6$ \cite{deltuva:18a}.
The results for 
$f\ge 5.3$ obtained solving the standard bound state FY equations
connect to $f\le 5.3$
results indicating the consistency between the simpler bound state and much
more complicated continuum calculations.
Surely, the resonance trajectory depends on the particular
enhancement scheme used and therefore is not identical with those
in Refs.~\cite{PhysRevC.93.044004,lazauskas:4n}. Nevertheless 
it exhibits a typical behavior
\cite{PhysRevC.93.044004,lazauskas:4n,lazauskas:3n,deltuva:18a}:
decreasing the enhancement factor $f$ the pole first moves to higher energy
and away from the real axis until the turning point where  
$E_r$ starts to decrease while $\Gamma$ continues increasing rapidly. 
For $f<4.3$ the pole of $\mct_{\beta\alpha}$ becomes
too far from the real axis  to be discernible from the
nonresonant continuum,
which results in  increasing theoretical error bars
estimated as in Ref.~\cite{deltuva:18a}.
Thus, an unrealistically large enhancement
of the higher-wave potential is needed to support an observable
$4n$ resonance, which strongly suggests that at the physical interaction
strength there is no observable $4n$ resonance, in agreement with
Refs.~\cite{PhysRevC.93.044004,lazauskas:4n}
and in contradiction with 
Refs.~\cite{PhysRevLett.117.182502,PhysRevLett.118.232501,PhysRevLett.119.032501}.

\begin{figure}[!]
\begin{center}
\includegraphics[scale=0.69]{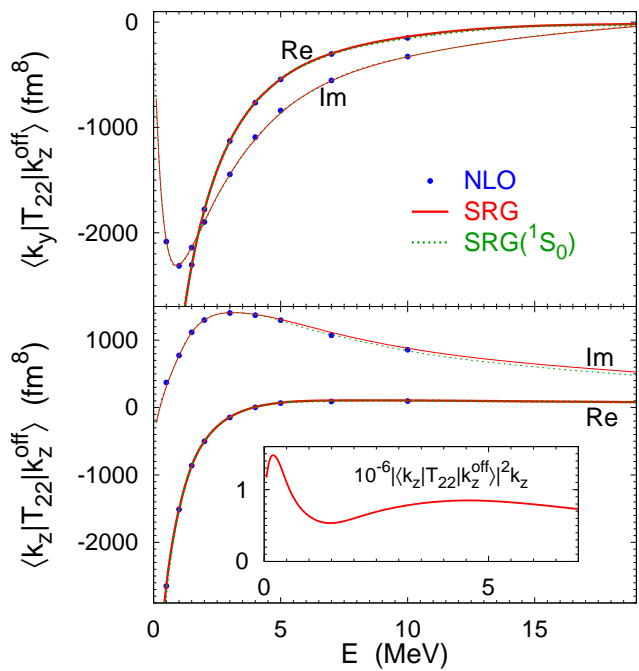}
\end{center}
\caption{\label{fig:tf1} (Color online)
Energy dependence of selected $4n$ transition matrix elements
obtained using the physical NLO (dots) and SRG (solid curves) potentials.
For the latter also the results including only the
${}^1S_0$  $nn$ interaction are given by dotted curves.
The inset shows the squared matrix element multiplied with 
 $k_z$ arising from the phase-space factor.
}
\end{figure}

The absence of an observable $4n$ resonance with a physical $nn$ interaction
is shown in Fig.~\ref{fig:tf1} over a broader energy range 
on the example of still another matrix elements of $\mct_{\beta\alpha}$.
Also predictions with  the NLO potential are presented. 
In fact, the results are almost independent 
of the force model, as observed also in previous works. 
Calculations using the CD Bonn potential
\cite{machleidt:01a}, not shown in Fig.~\ref{fig:tf1}, provide
an additional confirmation.
Furthermore, the results appear to be insensitive to 
$P-$ and higher-wave interaction: SRG calculations including only
the ${}^1S_0$ $nn$ force  agree quite well with full SRG results.
The dominance of  the $S$-wave interaction  may indicate
a manifestation of the four-fermion universality where
observables are governed by a large $nn$ scattering length.
This point of view also supports the absence of an
observable $4n$ resonance since the universal four-fermion system
is very far from being bound: 
 a positive scattering length for two difermions indicates that
their effective interaction 
is repulsive  \cite{deltuva:17d,petrov:04a}.


Despite that no observable $4n$ resonance is predicted, 
matrix elements of  transition operators  $\mct_{\beta\alpha}$ 
acquire large absolute values at low energies. This can be seen 
in both Figs.~\ref{fig:treim} and \ref{fig:tf1}, and is confirmed
by further calculations not shown here. One may conjecture that
this low-energy enhancement could manifest itself also in more
complicated many-body reactions with the $4n$ subsystem in the final
state such as  $\A{4}{He}(\A{8}{He},\A{8}{Be})$ 
of Ref.~\cite{PhysRevLett.116.052501}. The amplitude for such a reaction
could be approximated by a many-body double charge-exchange
matrix elements for the involved clusters ($\A{8}{Be}$ and $4n$)
 weighted with the corresponding initial and final-state
wave functions \cite{lazauskas:17a}. It also depends on the
double charge-exchange operator that is not well known;
note that a choice made in Ref.~\cite{lazauskas:17a}
has not produced a pronounced peak without a resonance.
Nevertheless, to illustrate the possibility of the
low-energy enhancement, in the inset of  Fig.~\ref{fig:tf1}
the squared matrix element of $\mct_{\beta\alpha}$
 multiplied with  $k_z$ due to the
phase-space factor is plotted; this product would be a factor
in the integrand determining 
 the cross section $d^6\sigma/d^3k_xd^3k_y$
for two (unbound) dineutrons. Indeed,
this quantity exhibits a two-peak shape: a sharp and narrow one
around 0.25 MeV and a broad one around 4.5 MeV.
Note, that peaks are possible even in repulsive systems;
a textbook example is given in Ref.~\cite{fluegge:book}.

Finally, it is important to understand the difference to
Refs.~\cite{PhysRevLett.117.182502,PhysRevLett.118.232501,PhysRevLett.119.032501}
that predicted a tetraneutron resonance. Among the approaches
used in those works there is also one  based on the 
 $nn$ force enhancement by a factor $f$ and subsequent extrapolation 
of the obtained bound-state energy to the $f=1$ limit in the 
continuum.\footnote{In Ref.~\cite{PhysRevLett.117.182502} this was an 
auxiliary method used beside the harmonic-oscillator representation.}
 However, 
Refs.~\cite{PhysRevLett.117.182502,PhysRevLett.118.232501,PhysRevLett.119.032501}
apply the same factor $f$
in all  $nn$ waves thereby generating a bound ${}^1S_0$  dineutron once
$f$ exceeds roughly 1.1. Thus, $4n$ states interpreted in
Refs.~\cite{PhysRevLett.117.182502,PhysRevLett.118.232501,PhysRevLett.119.032501}
as bound tetraneutrons are in fact above  the two-dineutron threshold.
 Strictly speaking, 
no stable $4n$ bound state is possible above the two-dineutron threshold,
only scattering states.  Thus, a calculation of $4n$ bound states
in the regime above the two-dineutron threshold and extrapolation
of their energies is meaningless.
A similar situation arises for the $4n$ system in an external trap 
where a  tetraneutron "bound" at the $4n$ threshold 
\cite{PhysRevLett.118.232501} 
is above the dineutron threshold.
These simple considerations indicate  serious shortcomings
in the calculations of 
Refs.~\cite{PhysRevLett.117.182502,PhysRevLett.118.232501,PhysRevLett.119.032501}
 and  question the reliability of their results.

\section{Conclusions \label{sec:sum}}

The $4n$ system was studied
using one of the most reliable four-nucleon continuum methods.
The integral equations for transition operators were 
solved in the momentum space leading to well-converged results.
Strongly enhancing the $nn$ force in higher partial waves 
the $4n$ model system in the $\mcj^\Pi =0^+$ state
was made bound or resonant.
In the latter case the resonant behavior was seen in 
all transition matrix elements, their energy dependence
was used to extract the resonance position and width.
However, reducing the enhancement factor the resonant behavior disappears 
well before the physical strength is reached. This indicates
the absence of an  observable $4n$ resonance,
in agreement with Refs.~\cite{PhysRevC.93.044004,lazauskas:4n}
and in contradiction with 
Refs.~\cite{PhysRevLett.117.182502,PhysRevLett.118.232501,PhysRevLett.119.032501}, 
a possible reason being the neglection of the dineutron threshold
in the latter works.

Even without an observable $4n$ resonance the
 transition operators exhibit pronounced low-energy peaks.
It is conjectured that they may be seen also in more
complicated reactions such as $\A{4}{He}(\A{8}{He},\A{8}{Be})$
of Ref.~\cite{PhysRevLett.116.052501} with the $4n$ subsystem
in the final state. The present calculation of half-shell matrix elements
of $4n$  transition operators is a first step toward 
understanding  of those reactions.

\vspace{1mm}

The author acknowledges discussions with R. Lazauskas,
the support  by the Alexander von Humboldt Foundation
under Grant No. LTU-1185721-HFST-E, and
the hospitality of the Ruhr-Universit\"at Bochum
where a part of this work was performed.


\end{document}